\documentclass[a4paper, 11pt]{article}    
\usepackage{latexsym}
\usepackage{amssymb}
\usepackage{amsmath}
\usepackage{amsfonts}
\usepackage{bbm}
\usepackage{graphicx}
\usepackage{float}
\usepackage[english]{babel}
\usepackage{multirow}
\usepackage{float}
\restylefloat{table}
\usepackage{caption}

\newcommand{\quota}[1]{``#1''}

\makeatletter         
\def\@maketitle{
\begin{center}
{\Huge \bfseries \sffamily \@title }\\[4ex] 
{\Large  \@author}\\[2ex] 
\includegraphics[width = 30mm]{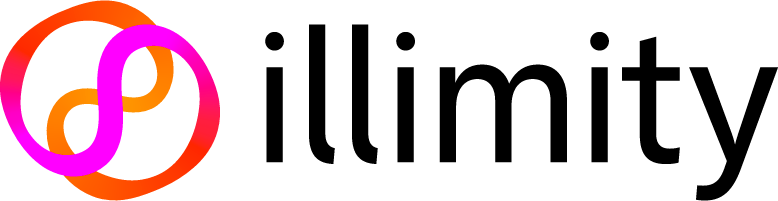}\\[4ex]
\@date\\[4ex]
\end{center}}
\makeatother

\title{ \bf A Simple Factoring Pricing Model}  
\author{Ilaria Nava\footnote{corresponding author: ilaria.nava@illimity.com}\,\,, Davide Cuccio, Lorenzo Giada and Claudio Nordio}

\begin{document}           
\maketitle                 

\begin{center}
\tt \Large Working paper\footnote{This paper reflects the authors' opinions and not necessarily those of their employers.}
\end{center}

\vspace{10mm}

\begin{abstract} 
\it \noindent
In a simplified setting, we show how to price invoice non-recourse factoring taking into account not only the credit worthiness of the debtor but also the assignor's one, together with the default correlation between the two. Indeed, the possible default of the assignor might impact the payoff by means of the bankruptcy revocatory, especially in case of undisclosed factoring. 
\end{abstract}

\bigskip

{\bf JEL} Classification codes: G12, G13, G33, G38 

{\bf AMS} Classification codes: 91B25, 91B70

\bigskip


\section{Standard Pricing Model}
\label{section:stdprimod}

\begin{figure}[!htb]
\centering
\caption{Schematic overview of a standard invoice factoring operation.}
\label{fig1}
\includegraphics[width=10cm]{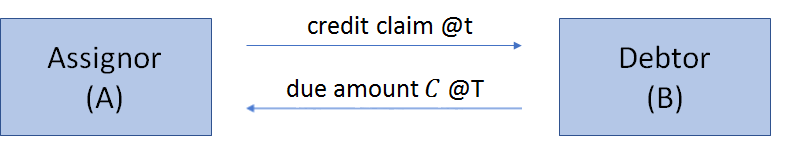} 
{\caption*{\it 1) Issuance of the receivable to the debtor and creation of the credit claim with maturity $T$. }}
\includegraphics[width=10cm]{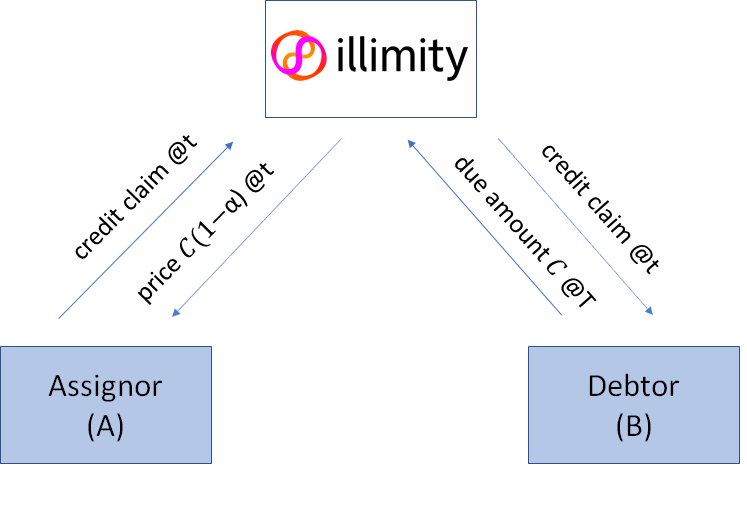}
{\caption*{\it 2) Assignment of the receivable to the factor upon payment of the purchase price $C(1-\alpha)$ and transfer of the enforcement power for the credit reimbursement to the assignee.}}
\end{figure}

Conventional practices in non-recourse factoring (see Figure \ref{fig1}) rely on the debtor's creditworthiness as a major determinant for pricing, or often consider it as the unique relevant parameter in the derivation of the expected payoff of an operation. In most cases, indeed, the default of the invoice debtor is assumed as the unique event with an actual impact on factoring proceedings. As a consequence, the standard pricing approach typically neglects  the occurrence of other circumstances which might condition the outcome of an operation, thus limiting the number of features considered in the computation of its fair value. 

According to this view, the problem of pricing in invoice factoring  can be simply represented by defining a set of complementary probabilistic events, which are centered on the conditions  deemed as crucial for the realization of the payoff.  
In the case of standard non-recourse factoring, therefore, such event corresponds to: 

\begin{equation}
    \mathbbm{1}_{\tau<T}=
    \begin{cases}
      1, & \text{if }\tau<T\\
      0, & \text{if }\tau\geqslant\ T
    \end{cases}
  \end{equation}
where ${\tau}$ is the time to default of the debtor, and $T$ the expected time of repayment of the receivable (i.e. the operation's maturity). When the default event occurs, i.e. $\mathbbm{1}_{\tau<T}=1$, the debtor defaults before repaying its debt, reducing the reimbursement obtainable by the assignee to a recovery value $rC$ (with $r$ indicating a recovery rate) typically smaller than the actual receivable value $C$. In the opposite case, the default of the debtor will occur after the credit maturity, allowing the assignee to fully receive the repayment of its credit. It follows, by definition, that the probability of the first event occurring is the default probability $PD_T$ of the debtor by time $T$, while the second event defines the probability of the debtor to survive until the conclusion of the factoring operation.

In symbols:
\begin{equation}\label{eq2}
\mathbb{E}\left(\mathbbm{1}_{\tau<T}\right) =\mathbb P(\tau<T)=PD_T.
\end{equation}
As we know different outcomes associated to each of the two events discussed, we can also define the overall payoff of the operation from the assignee's perspective as:
\begin{equation}\label{eq3}
\text{Payoff} = rC \cdot \mathbbm{1}_{\tau<T} + C\cdot \left(1-\mathbbm{1}_{\tau<T}\right),
\end{equation}
and its price - calculated as the expected payoff, not discounted for sake of simplicity - as
\begin{eqnarray}\label{eq3bis}
\text{Price} &=& rC \cdot \mathbb{E}\left(\mathbbm{1}_{\tau<T}\right) + C\cdot \left(1-\mathbb{E}\left(\mathbbm{1}_{\tau<T}\right) \right ) \nonumber \\
&=& rC \cdot \mathbb P(\tau<T) + C \cdot \left(1- \mathbb P(\tau<T) \right) = C \left( 1- (1-r) PD_T \right),
\end{eqnarray}
which depends exclusively on the default probability of the debtor, the recovery value in case of its default and the value of the outstanding invoice. The quantity $(1-r) PD_T$ must be smaller than the discount on the invoice purchase price $\alpha$, in order for the bank to make a profit, as depicted in the second part of Figure \ref{fig1}.

\section{The role of the assignor's default }


We now examine how the assignor's default enters into the standard pricing framework  presented in section \ref{section:stdprimod}, with the purpose to  better understand  its role compared to the one played by the debtor's default. 
 
The pricing problem for non recourse factoring should be adapted, in this case, to include a new set of events possibly affecting the outcome of an operation, each one depending on both the behaviour of the assignor and the debtor. 
According to the notation used so far, this new set is given by:

\begin{align*}
&Events:  &Payoffs:&\\
&\text{(a)} \hspace{10pt} \mathbbm{1}_{\tau_{A}< \tau_{B} < T} & r_{B}C\hspace{28pt}&\\[0.4em]
&\text{(b)} \hspace{10pt}\mathbbm{1}_{\tau_{B}< \tau_{A} < T} &  r_{B}C\hspace{28pt}&\\[0.4em]
&\text{(c)} \hspace{10pt} \mathbbm{1}_{\tau_{B}< T< \tau_{A}} & r_{B}C\hspace{28pt}&\\[0.4em]
&\text{(d)} \hspace{10pt} \mathbbm{1}_{T<\tau_{A}< \tau_{B}} &C \hspace{37pt}&\\[0.4em]
&\text{(e)} \hspace{10pt} \mathbbm{1}_{T<\tau_{B}< \tau_{A}} &C \hspace{37pt}&\\[0.4em]
&\text{(f)} \hspace{10pt} \mathbbm{1}_{\tau_{A}<T< \tau_{B}} &C \hspace{37pt}&\\[0.4em]
\end{align*}
where $A$ stays for the assignor, $ B$ for the debtor, and ${\tau_{A}}$,  ${\tau_{B}}$ for their respective times to default. The first event in the list, for instance,  corresponds to the case where both $A$, the assignor, and $B $,  the debtor,  default before the assignee manages to receive the payment of the purchased receivable at time $T$, with the assignor defaulting before its debtor. In this case the assignee will be able only to benefit from the recovery value associated to its credit towards the debtor, $r_{B}C$. Similarly, in the last event reported, the assignor $A$ will default before the maturity of the repayment, while the debtor $B$ will default only after having paid off its debt towards the assignee. This, of course, leads to a payoff for the assignee given by the full credit value, $C$. 
The probability of each of the events described above, therefore, can always be interpreted as the probability of two separate, elementary events occurring jointly. Taking event $(\text a)$  as an example, we can then recall the definition of joint probability as follows:

\begin{eqnarray}\label{eq4}
\mathbb{E}\left(\mathbbm{1}_{\tau_{A}< \tau_{B} <T}\right) &= & \mathbb P \left\{\left(\tau_{A}, \tau_{B} <T \right) \cap \left (\tau_{A}< \tau_{B} \right) \right\} \nonumber \\
&=&   \mathbb P\left(\tau_{A}, \tau_{B} <T | \tau_{A}< \tau_{B} \right) \cdot \mathbb P(\tau_{A}< \tau_{B}). 
\end{eqnarray}

By visually inspecting the list of the events above, it is easy to note that all the potential payoffs available to the assignee only depend on the time to default of the ceded debtor $\tau_{B}$, while no impact is actually due to to the potential default of the assignor. This means that the creditworthiness of the latter will not affect the pricing formula for factoring, which will predictably results to be equal to equation (\ref{eq3bis}). In fact, the expected payoff for the assignee can be synthesized using only two probabilistic events related to $\tau_{B}$, coming from the sum of events (a), (b), (c), and (d), (e), (f)  respectively. This leads to: 

\begin{equation}\label{eq5}
\text{Price} =  r_{B}C \cdot\mathbb{E} \left( \mathbbm{1}_{ \tau_{B} <T} \right ) +  C \cdot \left( 1- \mathbb{E} \left( \mathbbm{1}_ {\tau_{B}<T } \right)  \right)
\end{equation}
which, in probabilistic terms, is again equivalent to:

\begin{equation}\label{eq6}
\text{Price}=   r_{B}C \cdot \mathbb P(\tau_{B}<T) + C \cdot \left(1- \mathbb P(\tau_{B}<T) \right).
\end{equation}

It' s worth noting that this result holds for any value the default correlation between the two parties $A$ and $B$ might take, and regardless of any specific assumption about the shape of the marginal and  joint probability distributions of $\tau_{A}$ and $\tau_{B}$.
The result in (\ref{eq6}), however, should not lead to the conclusion that the creditworthiness of the assignor is always a negligible factor  in factoring pricing. In the following section, we will  indeed present a more refined approach where such variable comes into play, with a significant impact on the price determination. 

\section{Pricing under bankruptcy revocatory}

As it is the case for other financial transactions, even invoice factoring is generally exposed to the risk of the assignor's  bankruptcy by means of the application of the so called  claw-back actions (or bankruptcy revocatory).
Such procedures find their legal discipline within different frameworks of the Italian law. The Italian Bankruptcy Law\footnote{Art. 67, paragraphs I and II, specifies the type of transactions which are possibly subject to the application of claw back actions, and the related applicable suspect period. Among other requirements, the article states that the bankruptcy receiver also needs to prove that the other party was aware that the debtor was insolvent, upon finalization of the transaction (\quota{scientia decoctionis}).}, in the first place, provides that once the insolvency of a debtor has been declared all transactions effected by him over the previous year (or, in certain cases, over the previous six months) are scrutinized and possibly unwound as preferential. The major consequence is that the creditors' claims arisen in this period  are set aside until the completion  of the bankruptcy procedure and the distribution of the debtor' s assets among its creditors. A competitor  legal framework to this one, which specifically applies to the case of factoring,  is the Italian Factoring Law (Law of 21 February 1991, n. 52). This law establishes the unenforceability to third parties of the sale of a trade receivable  in case the finalization of the sale contract was made within a suspect period of one year prior to the bankruptcy of the seller\footnote{Art. 7 requires, similarly to the Bankruptcy Law, the existence of the \quota{scientia decoctionis} for the enforcement of the bankruptcy revocatory. This condition is typically verified in the case where the assignee of an invoice is a bank, which can check if the  debtor’s name is in the bulletin of unpaid negotiable instruments (e.g. notes, drafts, checks), or collect information about its solvency. }.
Despite relying on different assumption, rationales, and requirements, both these two legal regimes determine, where applicable, a unique outcome - the waiver of the factor's creditory claim towards the seller and the debtor - which can lead to a temporary or  permanent loss for the assignee in a factoring operation. This is all the more true in the case of  \quota{confidential} factoring, where the assignment of  the receivable has not been notified to the debtor, thus making the enforcement of the assignee' s claim even more difficult.

\begin{figure}[!htb]
\centering
\caption{Application of the assignor's bankruptcy revocatory after full repayment of the invoice amount, and subsequent implications on the assignee's outflows, as per events (m) or (n) above.}
\label{fig2}
\includegraphics[width=10cm]{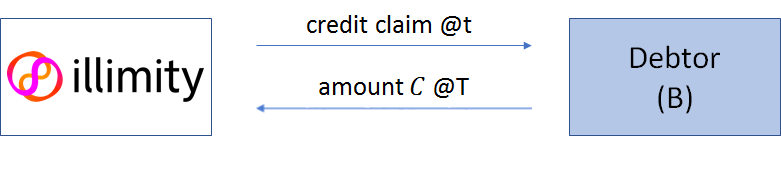}
{\caption*{\it 1) Payment by the debtor of the due receivable amount $C$ at the envisaged  maturity $T$.}}
\includegraphics[width=10cm]{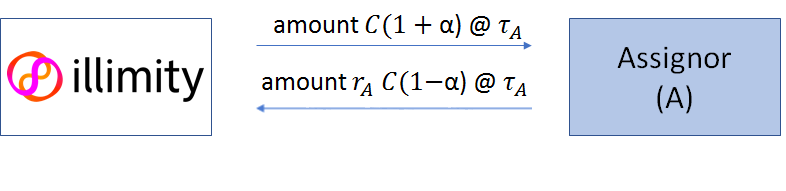}
{\caption*{\it 2) Default of the assignor in $\tau_{A}> T$  and return of the factoring proceedings $C$ collected by the assignee to the invoice seller, together with the fee gained on the credit, $\alpha C$. During the redistribution of the bankruptcy proceedings of the assignor, the assignee recovers a fraction of its financial exposure to the assignee, $r_{A}C\left( 1- \alpha\right)$.}}
\end{figure}

All this suggests that the bankruptcy revocatory of the assignor  might enter into the pricing procedure for factoring as a relevant event, adding to the debtor's default we  discussed in the previous sections. In this regard, we can define a new set of potential events impacting the payoff of factoring operations as follows: 

\begin{align*}
&\hspace{10pt} Events:  &Payoffs:\hspace{20pt}&\\
&\hspace{10pt} \text{(a)} \hspace{10pt}\mathbbm{1}_{\tau_{B}< \tau_{A}<T}\cdot \mathbbm{1}_{\tau_{A}> \Delta}  &  r_{B}C\hspace{43pt}&\\[0.4em]
&\hspace{10pt} \text{(b)} \hspace{10pt}\mathbbm{1}_{\tau_{A}< \tau_{B}<T}\cdot \mathbbm{1}_{\tau_{A}> \Delta} & r_{B}C\hspace{43pt}&\\[0.4em]
&\hspace{10pt} \text{(c)} \hspace{10pt}\mathbbm{1}_{\tau_{B}< T<\tau_{A}} \cdot \mathbbm{1}_{\tau_{A}> \Delta} & r_{B}C\hspace{43pt}&\\[0.4em]
&\hspace{10pt} \text{(d)} \hspace{10pt}\mathbbm{1}_{\tau_{A}< T<\tau_{B}}\cdot \mathbbm{1}_{\tau_{A}> \Delta} & C\hspace{56pt}&\\[0.4em]
&\hspace{10pt} \text{(e)} \hspace{10pt}\mathbbm{1}_{T<\tau_{B}<\tau_{A}}\cdot \mathbbm{1}_{\tau_{A}> \Delta} & C \hspace{56pt}&\\[0.4em]
&\hspace{10pt} \text{(f)} \hspace{10pt}\mathbbm{1}_{T<\tau_{A}<\tau_{B}}\cdot \mathbbm{1}_{\tau_{A}>\Delta}& C \hspace{56pt}&\\[0.4em]
&\hspace{10pt} \text{(g)} \hspace{10pt}\mathbbm{1}_{\tau_{B}< \tau_{A}<T}\cdot \mathbbm{1}_{\tau_{A}< \Delta} & - \alpha C + r_{A} C (1 - \alpha) \hspace{-28pt}&\\[0.4em]
&\hspace{10pt} \text{(h)} \hspace{10pt}\mathbbm{1}_{\tau_{A}< \tau_{B}<T}\cdot \mathbbm{1}_{\tau_{A}< \Delta} & - \alpha C + r_{A} C (1 - \alpha) \hspace{-28pt}&\\[0.4em]
&\hspace{10pt} \text{(i)} \hspace{10pt}\mathbbm{1}_{\tau_{B}<T< \tau_{A}}\cdot \mathbbm{1}_{\tau_{A}<\Delta} & - \alpha C + r_{A} C (1- \alpha) \hspace{-28pt}&\\[0.4em]
&\hspace{10pt} \text{(l)} \hspace{10pt}\mathbbm{1}_{\tau_{A}< T<\tau_{B}}\cdot \mathbbm{1}_{\tau_{A}< \Delta} &- \alpha C + r_{A} C (1 - \alpha) \hspace{-28pt}&\\[0.4em]
&\hspace{10pt} \text{(m)} \hspace{10pt}\mathbbm{1}_{T<\tau_{B}<\tau_{A}}\cdot \mathbbm{1}_{\tau_{A}< \Delta} & - \alpha C + r_{A} C (1 - \alpha) \hspace{-28pt}&\\[0.4em]
&\hspace{10pt} \text{(n)} \hspace{10pt}\mathbbm{1}_{T<\tau_{A}<\tau_{B}}\cdot \mathbbm{1}_{\tau_{A}<\Delta} &- \alpha C + r_{A}C (1 - \alpha) \hspace{-28pt}&\\[0.4em]
\end{align*}
where, on top of the notation used so far, we introduce the parameter $\Delta$, i.e. the suspect period for the enforcement of the bankruptcy revocatory in case of default of the assignor $A$. Note that, at this stage, we can assume that $\Delta$ features a suspect period of generic length, regardless of the specific judicial procedure that might be proposed by the bankruptcy administrator (Bankruptcy Law versus Factoring Law).
Furthermore, besides the payoffs $r_{B}C$ and $C$, we have now a third payoff, $-\alpha C + r_{A} C(1 - \alpha)$, characterizing the events where the regime of the revocatory applies  - see for instance Figure \ref{fig2} as a graphical representation of the case (m) or (n). $ 0<\alpha<1$, in the first term of this payoff, represents the discount charged  by the assignee on the nominal amount of the invoice upon its purchase, as discussed above. This quantity $\alpha C$ should be put back to the account seller in case of application of a clawback action. As mentioned, this always goes along with the return of the credit claim towards the ceded debtor to the assignor, or, in case the claim has already been cashed in, with the return of the proceeds collected  by the assignee until the default of the seller. The second term of the payoff reflects instead the positive recovery that the factor may obtain from the redistribution of the bankruptcy proceedings of A. As the assignee is entitled to lodge a claim equal to its original financial exposure to the defaulted assignor, this term is  expressed as $r_{A} C \left(1 - \alpha \right)$, where $0<r_{A}<1$  is the expected recovery percentage from the liquidation procedure of A, and $C (1-\alpha )$ the original exposure. 

It emerges, therefore, that the  payoff foreseen  in case of bankruptcy revocatory of the assignor is a function of the assignee's purchase price itself:
\begin{equation}\label{eq7.1}
\text{Price}= C \left( 1- \alpha \right). 
\end{equation}




We can sum up all the events associated to a certain payoff, in order to get a simpler view of the actual events in scope.
By following the same passages as in section 2, the assignee's expected payoff  boils down to:
\begin{equation}\label{eq8}
\text{Price}= r_{B}C \cdot \mathbb{E} \left( \mathbbm{1}_{\tau_{B}<T } \cdot \mathbbm{1}_{\tau_{A}> \Delta} \right ) +  C \cdot \mathbb{E} \left( \mathbbm{1}_{\tau_{B}>T } \cdot \mathbbm{1}_{\tau_{A}>\Delta} \right ) +( -\alpha C + r_{A}C \left( 1 - \alpha \right)) \cdot \mathbb{E} \left( \mathbbm{1}_{\tau_{A}< \Delta } \right)
\end{equation}\newline
which, in terms of the corresponding probabilities and of the relation in (\ref{eq7.1}), returns\footnote{For more details, see the appendix at section \ref{ap_1}} :\\
\begin{eqnarray}\label{eq9} 
\text{Price} &=&   r_{B}C  \cdot \mathbb P\left (\tau_{B}<T, \tau_{A}> \Delta \right) + C\cdot \mathbb P \left( \tau_{B}> T, \tau_{A} >\Delta \right)  + 
\left(  r_{A} C( 1 - \alpha) -\alpha C \right) \cdot \mathbb P\left(\tau_{A} < \Delta  \right) \nonumber \\
\nonumber \\
&=&  \frac{  r_{B}C  \cdot \mathbb P\left (\tau_{B}<T, \tau_{A}> \Delta \right) +  C\cdot \mathbb P \left( \tau_{B}> T, \tau_{A} >\Delta \right)  - C \cdot \mathbb P\left(\tau_{A} < \Delta  \right) }{1-\left(1+ r_{A} \right) \cdot \mathbb P \left(\tau_{A} < \Delta  \right)}.  
\end{eqnarray}

Equation (\ref{eq9}) shows therefore, how factoring operations could be priced when executed in a framework which envisions the application of revocatory actions in case of bankruptcy of the seller. The introduction of such regime makes the factor's payoff sensitive to the probability of default of its assignor, as it also appears by comparing equations (\ref{eq9}) and (\ref{eq6}). This implies that, ceteris paribus, the  lower the creditworthiness of the assignor, the higher its probability of defaulting within the bankruptcy suspect period, and thus the higher the probability of the assignee to incur a financial loss due to the revocatory procedure. As a consequence, the expected payoff  of the factor tends do decrease as the the default probability of the assignor increases. This effect is only mitigated by the possibility for the assignee to recover a fraction $r_{A}$ of its financial exposure in case of default of the seller, as  also indicated by the denominator of (\ref{eq9}). Vice versa, the higher the probability of the assignor not to default within such period, the higher the probability of the assignee to receive a higher positive payoff (the debtor's recovery rate $r_{B}C$ or the full credit amount $C$), jointly depending on the default probability of the debtor.
This suggests that the correlation between the time to default of the assignor and the debtor plays now a considerable role in the definition of the price. As it will be better illustrated in the next section, a larger correlation acts  by shifting the expected payoff towards the full recovery of the receivable amount.    

\section{Application}

In this section we will apply the pricing formula (\ref{eq9}) to show how it works in a simple case, where we define a specific distribution for the random variables $\tau_{A}$, $\tau_{B}$.
We assume, therefore, that the default times have a bivariate exponential distribution obtained combining exponential marginal distributions with constant default intensity\footnote{The constant default intensity $\lambda _{x}$ gives an exponential form to the survivalship probability $P (\tau_{x}>t)= e^{-\lambda _{x}t}$} with a Gumbel copula, disallowing simultaneous defaults. This particular choice of copula is common in the recent literature, and has the advantage of being analytically tractable. The resulting joint survival probability reads:

\begin{equation}\label{eq10}
\mathbb  P \left(\tau_{A}> t_{A} , \tau_{B} > t_{B} \right)= e ^{-[\left(\lambda_{A} t_{A}\right)^ {\theta} +\left(\lambda_{B} t_{B} \right) ^{\theta}]^{\frac{1}{\theta}}}
\end{equation}
with $\theta  \in [1, \infty)$, while the constant default intensity $\lambda _{x}$ gives an exponential form to the marginal survival probability $P (\tau_{x}>t)= e^{-\lambda _{x}t}$. 
Observe that Kendall’s Tau (a non-parametric measure of dependence between two random variables) $\tau_{K}=1- \frac{1}{\theta}$ , so that the independent case corresponds to $\theta=1$, and the comonotonic (fully dependent) case to $\theta=\infty$.
From eq.(\ref{eq10}), the joint cumulative probability function of the default times is:
\begin{equation}\label{eq11}
F_{A,B}(t_{A}, t_{B}) =\mathbb  P \left(\tau_{A}<t_{A} , \tau_{B} < t_{B} \right)= e ^{-[\left(\lambda_{A} t_{A}\right)^ {\theta} +\left(\lambda_{B} t_{B} \right) ^{\theta}]^{\frac{1}{\theta}}} - e^{-\lambda_{A}t_{A}} +1 -e^{-\lambda_{B}t_{B}}
\end{equation}
and the joint probability density function is:
\begin{eqnarray}\label{eq12}
f_{A,B}(\tau_{A}, \tau_{B}) &=& \left[ \theta  -1 + \left(\lambda_{A}^{\theta}\tau_{A}^{\theta} +{\lambda}_{B}^{\theta}\tau_{B} ^{\theta}\right)^\frac{1}{\theta} \right] 
\left(\lambda_{A}^{\theta} \tau_{A}^{\theta}+ \lambda_{B}^{\theta}\tau_{B}^{\theta} \right)^{\left(\frac{1}{\theta}-2\right)} \cdot \nonumber \\
&  & \lambda_{A}^{\theta} \lambda_{B}^{\theta}\tau_{A}^{\theta -1} \tau_{B}^{\theta -1} e ^{-[\left(\lambda_{A} \tau_{A}\right)^ {\theta} +\left(\lambda_{B} \tau_{B} \right) ^{\theta}]^{\frac{1}{\theta}}}.
\end{eqnarray}

Having set this out, we can simply convert the probabilities featuring the pricing equation (\ref{eq9}) in terms of the specified joint distribution as:
\begin{align}
\mathbb P\left (\tau_{B}<T, \tau_{A}> \Delta \right) &= \int_{\Delta}^{\infty}  d\tau_{A} \int_{0}^{T} f_{A,B}(\tau_{A}, \tau_{B} ) d\tau_{B} \label{eq_a} \\
\mathbb P \left( \tau_{B}> T, \tau_{A} >\Delta \right) &= \int_{\Delta}^{\infty} d \tau_{A} \int_{T}^{\infty} f_{A,B}(\tau_{A}, \tau_{B} ) d\tau_{B} \label{eq_b}\\                                           
\mathbb P\left(\tau_{A} < \Delta  \right) &= \int_{0}^{\infty} d\tau_{B} \int_{0}^{\Delta} f_{A,B}(\tau_{A}, \tau_{B} ) d\tau_{A} \label{eq_c} 
\end{align}
Thus, the overall is results equal to\footnote{For more details, see the appendix at section \ref{ap_2}. }:

\begin{eqnarray}\label{eq13}
\text{Price} &=&\frac{ r_{B}C \left( - e ^{-[\left(\lambda_{A} \Delta\right)^ {\theta} +\left(\lambda_{B} T \right) ^{\theta}]^{\frac{1}{\theta}}} + e^{-\lambda_{A} \Delta} \right) + C \left(  e ^{-[\left(\lambda_{A} \Delta\right)^ {\theta} +\left(\lambda_{B} T \right) ^{\theta}]^{\frac{1}{\theta}}} \right) - C \left( 1- e^{- \lambda_{A}\Delta }\right)}{e^{- \lambda_{A} \Delta}- r_{A} \left(1-e^{- \lambda_{A} \Delta} \right)} \nonumber\\
& = &   C\frac{ \left(1-r_B\right) e ^{-[\left(\lambda_{A} \Delta\right)^ {\theta} +\left(\lambda_{B} T \right) ^{\theta}]^{\frac{1}{\theta}} }   -1 + e^{-\lambda_{A} \Delta}  \left( 1+r_B\right)}
{e^{- \lambda_{A} \Delta} \left(1+r_A \right)- r_{A} } .
\end{eqnarray}

This formulation provides a better insight on how the correlation parameter $\theta$ enters into the definition of the price. An increase of the default correlation will cause a reduction of the probability of the factor to get the recovery payoff $r_{B}C$, i.e. the probability that debtor defaults before the repayment period while no revocatory action is enforceable towards the assignee (eq.~(\ref{eq_a})). On the other hand, a higher default correlation also implies a higher probability that both the debtor and the assignor will jointly survive until the maturity of the operation, allowing the assignee to receive the full reimbursement of its credit (eq. (\ref{eq_b})). This second effect always prevails on the first one, thus making the expected payoff of the assignee an increasing function of $\theta$.
Furthermore, as stated before, the payoff will discount a higher reduction as the probability of default of the assignor increases (eq. (\ref{eq_c})), other things being unchanged\footnote{These results are obviously influenced by the specific choice of copula we used in this example, which implies a symmetric correlation between $\tau_{A}$ and $\tau_{B} $. Alternative copula models, as those implying a skewed correlation between the times to default, could be a valid substitute approach to the one discussed, and lead to different results, for example by increasing the correlation between default, without modifying that between survival. }.  

Below, we show these results assuming a suspect period of half the length of the factoring maturity ($ \Delta=0.5$ and $T=1$), and an invoice nominal amount $C=100$. For simplicity, we assume that the recovery rates from the default of  A and B are both equal to $20\%$, so that $r_{A}=r_{B}=20\%$. The expected payoff of the assignee is first computed according to the conventional pricing formula in (\ref{eq6}), i.e. considering only the debtor's default probability. We assume that the debtor's default intensity is $\lambda_{B}=0.1$, which corresponds to a $PD_{B}=9.5\%$ - in line with a single B rating -  and a credit spread of 7.6\% on an ordinary loan. The pricing equation (\ref{eq13}) is then applied for comparison purposes, firstly taking the default intensity of the assignor as $\lambda_{A}=0.1$, and secondly as  $\lambda_{A}=0.2$, corresponding to a credit spread twice as large as the first one.\\


\begin {table}[H]
\centering
  \begin{tabular}{|c|c|c|c|c|}
   \hline
    \multirow{2}{*}{$\theta$ }& \multirow{2}{*}{ $\tau_K$} & \multicolumn{3}{|c|}{Prices}\\
    & & $\lambda_{B}$=0.1 &  $\lambda_{B}$=0.1, $\lambda_{A}$=0.1 &   $\lambda_{B}$=0.1, $\lambda_{A}$=0.2 \\
   \hline
    1     & 0.00  & 92.387 & 88.164 & 83.629 \\
    2     & 0.50  & 92.387 & 91.011 & 88.089 \\
    3     & 0.67  & 92.387 & 91.606 & 89.309 \\
    4     & 0.75  & 92.387 & 91.796 & 89.873 \\
    5     & 0.80  & 92.387 & 91.866 & 90.199 \\
   \hline
\end{tabular}
\caption{Application of pricing formulas (\ref{eq6}) and (\ref{eq13}) for different values of $\theta$, and default intensities $\lambda_{A}$, $ \lambda_{B}$; $ \Delta=0.5$ and $T=1$. }
\label{table1}
\end{table}

In the next table we show the results when the time to repayment of the loan $T = 0.5$ is shorter than the suspect period $\Delta = 1$. In this case, the default probability of the debtor is obviously about half that of the previous example, and both the first to default effect and the correlation effect are even more pronounced. This is quite a significant result, as one might think that even if the assignor is quite risky, the transaction profitability depends only on the credit quality of the debtor, while one can see that the main contribution comes from the default probability of the assignor.
\begin {table}[H]
\centering
  \begin{tabular}{|c|c|c|c|c|}
   \hline
    \multirow{2}{*}{$\theta$ }& \multirow{2}{*}{ $\tau_K$} & \multicolumn{3}{|c|}{Prices}\\
    & & $\lambda_{B}$=0.1 &  $\lambda_{B}$=0.1, $\lambda_{A}$=0.1 &   $\lambda_{B}$=0.1, $\lambda_{A}$=0.2 \\
   \hline
    1     & 0.00  & 96.09835 & 87.42007 & 77.38472 \\
    2     & 0.50  & 96.09835 & 90.44666 & 80.95348 \\
    3     & 0.67  & 96.09835 & 91.07899 & 81.38043 \\
    4     & 0.75  & 96.09835 & 91.28085 & 81.45081 \\
    5     & 0.80  & 96.09835 & 91.35512 & 81.46387 \\
   \hline
\end{tabular}
\caption{Application of pricing formulas (\ref{eq6}) and (\ref{eq13}) for different values of $\theta$, and default intensities $\lambda_{A}$, $ \lambda_{B}$; $ \Delta=1$ and $T=0.5$. }
\label{table1}
\end{table}

\section{Conclusions}


Assessing the creditworthiness of a debtor is a key element for non-recourse invoice factoring, as it affects the evaluation by factoring facilities of the actual profitability of such operations. As a matter of fact, the standard  pricing techniques in this field often focus on the debtor's default as the only relevant event for price determination, while neglecting the potential  role played by the default of the receivable assignor, or its correlation with the debtor's one. These events, in fact, turn out to be  relevant whenever they exert a direct binding effect on the payoff of the factor, as it happens, for instance, by means of the application of the bankruptcy revocatory procedure. Under this regime, the factor credit claim towards the debtor, or the proceedings already collected from the latter, are put back to the receivable seller in case of declaration of its bankrupt, provided that the receivable sale contract was finalized in a period of time (usually six months to one year) prior to the declaration of default. This determines, especially in the case of non notification factoring, the inability of the assignee to enforce its position towards the ceded debtor. Furthermore, it implies the obligation for the assignor to return the invoice discount originally retained upon purchase and thus a potential financial loss at the end of the revocatory procedure. The larger the assignor' s probability of default, therefore, the higher the probability for the assignee to fall into such revocatory framework, which translates, eventually, into a lower expected payoff. In addition to this, the assignee's return does also depend on the joint riskiness of the assignor and the debtor, as the relative timing of their default events can substantially influence the payoff of the operation. In particular, the higher the default correlation between the two, the higher, ceteris paribus, their probability  to jointly survive until the time due for the repayment of the receivable, allowing the assignee to receive the value of the outstanding invoice. These considerations  feature therefore a different pricing approach than the conventional one, as it admits, under bankruptcy revocatory, the inclusion of the creditworthiness of the assignor and its correlation with the debtor's default  as additional determinants of factoring prices.


\section*{Acknowledgments}
We are grateful to Corrado Passera for encouraging our research.

\appendix
\section{Appendix}
\subsection{Derivation of equation (\ref{eq9})}\label{ap_1}
We briefly derive in this section the pricing equation (\ref{eq9}), by substituting the relation in (\ref{eq8}) into the assignee' s expected payoff:
\begin{equation}
\text{Price} =   r_{B}C  \cdot \mathbb P\left (\tau_{B}<T, \tau_{A}> \Delta \right) + C\cdot \mathbb P \left( \tau_{B}> T, \tau_{A} >\Delta \right)  +\left( -\alpha C +  r_{A} C( 1 - \alpha) \right) \cdot \mathbb P\left(\tau_{A} < \Delta  \right). \nonumber \\
\end{equation}

As both the discount $\alpha C$ and the financial exposure of the assignee $C(1- \alpha)$ are  clearly  a function of the purchase price of the invoice, it is necessary to make the latter explicit into the  above equation.
Hence, replacing equation (\ref{eq8}) into the previous one, we obtain:

\begin{equation}
\text{Price}=  r_{B}C  \cdot \mathbb P\left (\tau_{B}<T, \tau_{A}> \Delta \right) + C\cdot \mathbb P \left( \tau_{B}> T, \tau_{A} >\Delta \right)+ \left( \text{Price}- C +r_{A}\text{Price}\right) \cdot P\left(\tau_{A} < \Delta  \right) \nonumber
\end{equation}
which, factoring out the price variable, returns:
\begin{equation}
\text{Price} \left(1- (1+ r_{A})\cdot P\left(\tau_{A} < \Delta \right) \right)=  r_{B}C  \cdot \mathbb P\left (\tau_{B}<T, \tau_{A}> \Delta \right) + C\cdot \mathbb P \left( \tau_{B}> T, \tau_{A} >\Delta \right)  - C \cdot \mathbb P\left(\tau_{A} < \Delta  \right) \nonumber
\end{equation}
and eventually:
\begin{equation}
\text{Price}= \frac{  r_{B}C  \cdot \mathbb P\left (\tau_{B}<T, \tau_{A}> \Delta \right) +  C\cdot \mathbb P \left( \tau_{B}> T, \tau_{A} >\Delta \right)  - C \cdot \mathbb P\left(\tau_{A} < \Delta  \right) }{1-\left(1+ r_{A} \right) \cdot \mathbb P \left(\tau_{A} < \Delta  \right)}.    
\end{equation}

\subsection{Derivation of equation (\ref{eq13})}\label{ap_2}
As stated in section 4, equation (\ref{eq13}) is obtained as the sum of the three integrals in equations (\ref{eq_a}) - (\ref{eq_c})  multiplied by the respective payoffs.\\
As for equation (\ref{eq_a}) one has:
\begin{align}
\mathbb P\left (\tau_{B}<T, \tau_{A}> \Delta \right) &= \int_{\Delta}^{\infty}  d\tau_{A} \int_{0}^{T} f_{A,B}(\tau_{A}, \tau_{B} )  d\tau_{B}. \nonumber \\
\end{align}

Making use of the well-known definition of the marginal and joint cumulative functions of $\tau_{A}, \tau_{B}$, i.e. $F_{A}(\tau_{A})$, $F_{B}(\tau_{B})$ and $F_{A, B}(\tau_{A},\tau_{B})$, we can more easily compute the above expression as:
\begin{align}
\mathbb P\left (\tau_{B}<T, \tau_{A}> \Delta \right) &= F_B(T) - F_{AB}\left(\tau_B=T,\tau_A=\Delta\right) \nonumber \\
							       &= \left ( 1-e^{-\lambda_{B} T} \right ) -\left(e ^{-[\left(\lambda_{A} \Delta \right)^ {\theta} +\left(\lambda_{B} T\right) ^{\theta}]^{\frac{1}{\theta}}}-e^{-\lambda_{A}\Delta}+1-e^{-\lambda_{B} T}\right) \nonumber \\
&= -e ^{-[\left(\lambda_{A} \Delta \right)^ {\theta} +\left(\lambda_{B} T \right) ^{\theta}]^{\frac{1}{\theta}}} + e^{-\lambda_A\Delta}\label{eq_abis}.
\end{align}
Accordingly, equation (\ref{eq_b}) gives:  
\begin{align}
\mathbb P \left( \tau_{B}> T, \tau_{A} >\Delta \right) &= \int_{\Delta}^{\infty} d \tau_{A} \int_{T}^{\infty} f_{A,B}(\tau_{A}, \tau_{B} ) d\tau_{B} \nonumber \\
							   &=1-F_A(\Delta)-F_B(T)+F_{AB}(\tau_{B}=T, \tau_{A}=\Delta)  \nonumber \\
							   &=1-\left(1-e^{-\lambda_A\Delta}\right)-\left(1-e^{-\lambda_BT}\right)+e ^{-[\left(\lambda_{A} \Delta \right)^ {\theta} +\left(\lambda_{B} T \right) ^{\theta}]^{\frac{1}{\theta}}}-e^{-\lambda_{A}\Delta}+1-e^{-\lambda_{B} T}\nonumber \\
							   &=e ^{-[\left(\lambda_{A} \Delta \right)^ {\theta} +\left(\lambda_{B} T \right) ^{\theta}]^{\frac{1}{\theta}}} \label{eq_bbis},
\end{align}
which nicely agrees with the result reported in equation (\ref{eq10}).\\
Equation (\ref{eq_c}) is trivially:
\begin{align}
\mathbb P\left(\tau_{A} < \Delta  \right) &= \int_{0}^{\infty} d \tau_{B} \int_{0}^{\Delta} f_{A,B}(\tau_{A}, \tau_{B} ) d\tau_{A}\nonumber \\
                                                                                                      &=F_{A}(\Delta) \nonumber \\
									      &=1-e^{-\lambda_A\Delta} \label{eq_cbis}
\end{align}
By summing up the contributions in equations (\ref{eq_abis}) - (\ref{eq_cbis}) one immediately obtains the result in equation (\ref{eq13}).


\begin{thebibliography}{99}

\bibitem{shortdesofref} \textsc{L. Giada, C. Nordio}, {\it Breaking break clauses}, RISK Magazine, March 2013
\bibitem{shortdesofref} \textsc{A. McNeil, R. Frey, P. Embrechts}, {\it Quantitative Risk Management }, Princeton University Press, 2005
\bibitem{shortdesofref} \textsc{EU Federation for Factoring and Commercial Finance}, {\it Factoring and Commercial Finance: An Introduction }, 2017
\end{thebibliography}
\end{document}